\documentclass[twoside]{article}
\usepackage{fleqn,espcrc2}
\usepackage{graphicx,times}
\usepackage{amssymb}
\begin{document}
\title{Origin of the pseudogap phase: Precursor superconductivity 
versus a competing energy gap scenario}

\author{K. Levin$^1$,
Qijin Chen$^2$,
Ioan Kosztin$^3$,
Boldizs\'ar Jank\'o$^4$,
Ying-Jer Kao$^1$, and
Andrew Iyengar$^1$\\\vspace*{1.5ex}
\textit{$^1$James Franck Institute, University of Chicago, 5640 South
  Ellis Avenue, Chicago, Illinois 60637, USA\\
$^2$National High Magnetic Field Lab, Tallahassee, Florida 32310, USA\\
$^3$Beckman Institute and Department of Physics, University of
     Illinois, Urbana, Illinois 61801, USA\\
$^4$Department of Physics, University of Notre Dame, Notre
  Dame, Indiana 46556, USA}}

\begin{abstract}
  
  In the last few years evidence has been accumulating that there are a
  multiplicity of energy scales which characterize superconductivity in
  the underdoped cuprates. In contrast to the situation in BCS
  superconductors, the phase coherence temperature $T_c$ is different
  from the energy gap onset temperature $ T^*$. In addition,
  thermodynamic and tunneling spectroscopies have led to the inference
  that the order parameter $\Delta_{sc}$ is to be distinguished from the
  excitation gap $\Delta$; in this way, pseudogap effects persist below
  $T_c$.  It has been argued by many in the community that the presence
  of these distinct energy scales demonstrates that the pseudogap is
  unrelated to superconductivity.  In this paper we show that this
  inference is incorrect. We demonstrate that the difference between the
  order parameter and excitation gap and the contrasting dependences of
  $T^*$ and $T_c$ on hole concentration $x$ and magnetic field $H$
  follow from a natural generalization of BCS theory.  This simple
  generalized form is based on a BCS-like ground state, but with self
  consistently determined chemical potential in the presence of
  arbitrary attractive coupling $g$. We have applied this mean field
  theory with some success to tunneling, transport, thermodynamics and
  magnetic field effects.  We contrast the present approach with the
  phase fluctuation scenario and discuss key features which might
  distinguish our precursor superconductivity picture from that
  involving a competing order parameter.
\hfill \textsf{\textbf{cond-mat/0107275}}

\end{abstract}

%
%



\maketitle


One of the biggest questions which faces the high temperature
superconductivity community is determining the origin of the pseudogap
phase. It is now becoming clear that pseudogap effects are not
restricted to the normal state alone. Moreover, they appear to persist
over a wide range of the phase diagram, up to and possibly above optimal
doping\cite{Timuskreview}. Understanding the pseudogap phase is
essential in order to find a proper replacement for BCS theory.  Indeed,
the failure of BCS theory is demonstrated most clearly in
thermodynamical\cite{Loram} and tunneling\cite{Renner} data which have
made it clear that the underlying normal phase \textit{below $T_c$}
contains a (pseudo)gap in the excitation spectrum. The phase diagram,
itself, indicates that the larger the pseudogap the lower is $T_c$ and
in this way the pseudogap appears to compete with superconductivity.
This competition has led many researchers to argue that $ T _c$ and $T
^* $ must necessarily originate from different physical mechanisms. A
recent D- density wave (DDW) theory\cite{Laughlin} presents a concrete
realization of the ``competing energy gap" scenario, first conjectured
by Loram and co-workers\cite{Loram}.

The present paper summarizes work from our
group\cite{ourpapers,Kosztin1,Chen1,Chen2,Chen3,ThermoPRB} which is
based on a precursor superconductivity approach to understanding the
pseudogap phase. We have been arguing for some time now that pseudogap
effects necessarily persist below $T_c$, and moreover, appear to compete
with superconductivity -- despite their common origin.  Our approach is
based on a simple physical picture\cite{Uemura} which interpolates
smoothly between BCS theory and Bose Einstein condensation, and on its
realization as first studied by Leggett\cite{Leggett},
Nozieres\cite{NSR} and Randeria\cite{Randeriareview} and their
respective co-workers. Our main contribution has been to extend the
Leggett ground state picture to non zero temperatures in a self
consistent fashion, so as to incorporate pseudogap effects which were
absent in earlier finite temperature calculations. This is a manifestly
mean field approach which should be contrasted with the phase
fluctuation scenario\cite{Emery}. The latter builds on strict BCS theory
(which is a special case of our more general theory), but goes beyond to
include fluctuations of the order parameter phase, which we neglect
here.  It should also be stressed that a strong case can be made for
focusing as we do here on an alternative mean field theory, as
distinguished from a fluctuation scheme. A nice summary of the
experimental support for this viewpoint, (principally based on the
observed narrow fluctuation regime), can be found in Ref.~\cite{Larkin}.

The key assumption underlying our theoretical approach is that the
ground state is of the generalized BCS form with wave-function
\begin{equation}
\Psi_0 = \prod_{\bf k} [ u_{\bf k} + 
v_{\bf k} c_{\bf k} ^ \dagger c_{-{\bf k}} ^\dagger ] |0\rangle
\end{equation}
%
Here, however, the chemical potential $\mu$ is determined self
consistently with variable attractive coupling constant $g$. As $g$
increases, $\mu$ progressively decreases, ultimately becoming negative
in the ``bosonic" regime. In this approach it should be clear that
superconductivity need not be associated with an underlying Fermi liquid
state. \textit{Without doing any calculations} one can make a number of
inferences about the finite temperature behavior associated with Eq(1).
It is clear that at sufficiently strong coupling, the onset for pair
formation $T^*$ will be different from $T_c$\cite{NSR}. It then
follows\cite{Kosztin1} that the excitation gap $\Delta$ and mean field
order parameter $\Delta_{sc}$ are necessarily distinct-- except,
however, at zero temperature, where there is full condensation of all
pairs ($\Delta = \Delta_{sc}$), as is implicit in Eq.~(1).

It should not be surprising, given the BCS-like form of the ground state
wave-function, that one arrives at self consistent equations for
$\Delta$ and the chemical potential $\mu$ which are essentially those of
BCS theory.  Here, however, the dispersion relation for fermionic
excitations is
\begin{equation}
E_{\bf k} = \sqrt{ \Delta ^2 \varphi_{\bf k}^2 + 
( \xi_{\bf k} - \mu ) ^2 }, \,\,\,\, 
\Delta ^2 = \Delta_{sc} ^2 + \Delta_{pg} ^ 2 
\end{equation}
%
where $\varphi_{\bf k}$ represents $s$- or $d$-wave
symmetry and the essential deviation from BCS theory derives from the
presence of the pseudogap $\Delta_{pg}(T) $ which must be determined
from a third self consistency condition (not given here). This latter
equation, in effect, differentiates our approach from
others\cite{Loram,Laughlin} which also have a BCS-like structure, but in
which $\Delta_{pg}$ is presumed to arise from a non-pairing channel.
Here, by contrast, the pseudogap arises from the stronger (than BCS)
attractive interaction, and is associated with \textit{excitations} of
finite momentum pairs. The underlying microscopic theory behind our
approach is based on a T-matrix decoupling scheme of Kadanoff and
Martin\cite{Kadanoff} which is discussed elsewhere\cite{Chen3}.  This
T-matrix can be thought of as a propagator for pairs which, in turn,
depends on the single particle self energy. The latter, in turn, depends
on the T-matrix-- so that particles and pairs interact and no higher
order correlation functions are included.  Stated simply, whereas BCS is
a mean field treatment of the fermions (and condensate), here we go
beyond BCS to include non-condensed (and essentially non-interacting)
pairs which are correspondingly treated at the mean field level.

With this well defined many body scheme, one can compute two particle
properties such as as the ac conductivity and superfluid density
$\rho_s$\cite{Chen1}.  The conductivity diagrams are related to the well
known Maki-Thompson and Aslamazov-Larkin terms (but with additional self
energy insertions\cite{Patton}).  A relatively simple expression for
$\rho_s$ results, which is of the BCS form, but with a prefactor of
$\Delta_{sc}^2 (T) $, where everywhere else in the expression, the full
energy gap $\Delta$ appears. What needs to be stressed here is that
$\rho_s$ reaches zero (at $T_c$) ``prematurely" before the fermionic
excitation gap $\Delta$ vanishes and that this occurs because there are
bosonic degrees of freedom (finite momentum pair excitations) in
addition to the usual $d$-wave fermions.  The presence of additional
bosonic excitations is, perhaps, the simplest way to understand the
experimental observation that $\rho_s$ scales with $T/T_c(x)$ rather
than with $T / \Delta(x)$.

 \begin{figure}
   \centerline{ \includegraphics[width=2.8in]{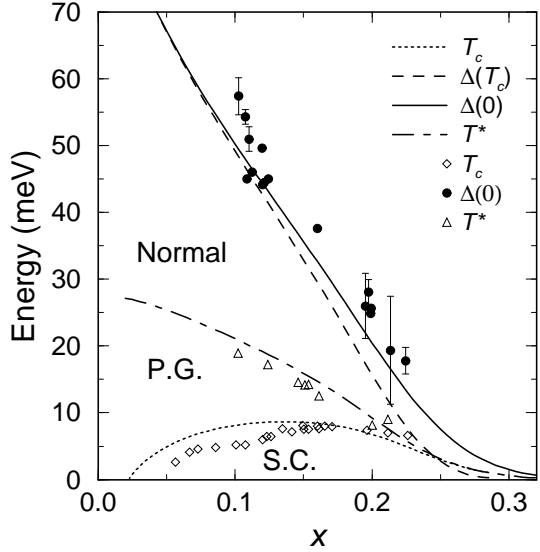} } 
   \vskip -1.1cm
     \caption{\small Comparison between theoretically computed 
       (with one fitting parameter) and measured phase diagram. 
     Experimental data are taken from: ($\bullet$) Ref.~\cite{JohnZ};
     ($\diamond$) Ref.~\cite{Rossat-Mignod}; ({\scriptsize $\triangle$})
     Ref.~\cite{Oda}. For details see Ref.~\protect\cite{Chen1}.}
   \vspace*{-4ex}
 \end{figure}  

In order to pass from coupling constant $g$ to hole concentration $x$,
we note that the effective coupling (which always enters in a
dimensionless form as a ratio to the bandwidth) is increased with
underdoping, since electronic energy scales decrease as the Mott
insulator is approached.  Assuming $g$ is $x$ independent, and with one
free parameter\cite{Chen1}, we find a reasonable fit to the phase
diagram as shown in Fig.~1. Indeed, the measured ``universality" in a
plot of $\rho_s(T,x)/\rho_s(0,x)$ vs $ T / T_c(x)$ appears to be rather
well satisfied within our theory, as is shown in Fig.~2 below-- along
with a comparison of data from Ref.\cite{Panagopoulos}. It should be
stressed that this methodology for incorporating Mott insulating physics
can be viewed as a device for fitting the phase diagram and is not an
essential component of the good agreement between theory and experiment
shown in Fig.~2.

 \begin{figure}[tbh]
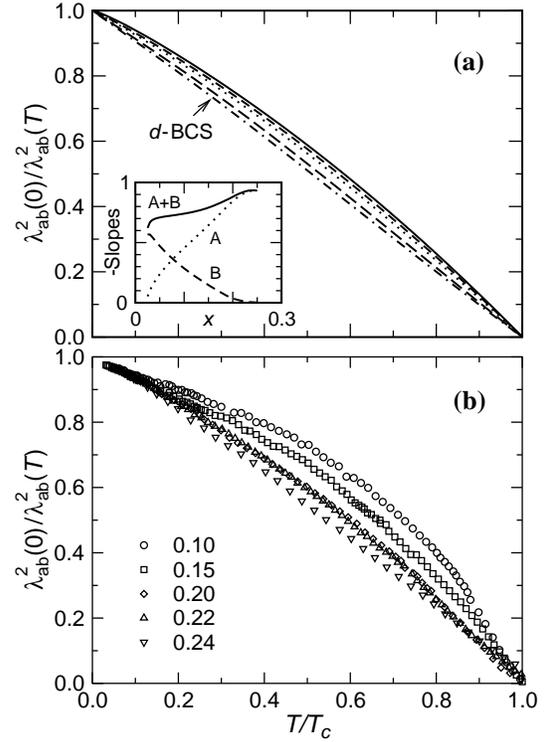

   \centerline{ 
\includegraphics[bb = 189 179 563 423, width=2.75in,clip]{Fig2a.eps} }
   \centerline{ \includegraphics[width=2.75in,clip]{Fig2b.eps} } 
   \vskip -1cm
   \caption{\small (a) Theoretical predictions for and (b) experimental
    measurements of the temperature dependence of the ab-plane inverse
    squared penetration depth.  Systematic trends with hole
    concentration are similar.  In (a), from bottom to top: $x=0.25$
    (BCS limit, dot-dashed line), 0.2 (long-dashed), 0.155 (dotted),
    0.125 (dashed) and 0.05 (solid line). At low $T$,
    $\lambda^2_L(0)/\lambda^2_L(T) = 1-[A+B(T)](T/T_c)$, where
    $B(T)\propto \sqrt{T/T_c}$ depends on $T$ very weakly. Shown in the
    inset are the low temperature values for A and B.  Experimental data
    on LSCO shown in (b) are taken from
    Ref.~\protect\cite{Panagopoulos}.  }
   \vspace*{-4ex}
 \end{figure}  

The bosons in the present theory, have a quasi-ideal gas character, as
might be expected based on the fully condensed ground state
wavefunction, and this leads naturally to a low temperature rounding
($T^{3/2}$, which enters with $\Delta_{sc}^2$) of the inverse square of
the penetration depth.  Indeed, it is difficult to find samples where
there is no low $T$ deviation from the expected $d$-wave linear
dependence,\cite{Lemberger} although this rounding has generally been
attributed to impurity effects (and fitted to $T^2$).  It should be
stressed that the bosonic degrees of freedom will show up quite
generally in the finite frequency conductivity as well.

The presence of a pseudogap in the fermionic spectrum at $T_c$ has a
number of important consequences. Because of the BCS like structure of
the equation for $\Delta$, it follows that $T_c$ will be suppressed as
$\Delta_{pg}$ increases.  Physically, this is a consequence of the
depressed density of states which makes it difficult for fermions to
pair.  This, in turn, is consistent with the general trends in the phase
diagram with increased underdoping.  The presence of this gap is also
responsible for the stronger sensitivity to pair breaking perturbations,
\textit{i.e.} magnetic fields\cite{companionpaper} and impurity
substitutions, seen in $T_c$, but not in $T^*$.  This latter observation
should counter the wide-spread inference that the contrasting behavior
of $T_c$ and $T^*$ with respect to $H$ and other perturbations is
suggestive of a competing rather than precursor origin of the pseudogap.
Finally, the presence of a pseudogap at $T_c$ raises the question of
what is responsible for the signature of true phase coherence at $T_c$,
given that an excitation gap is present when superconductivity is
established.  Our recent work\cite{ThermoPRB} has addressed this
question in more quantitative detail. One can view the excitations of
the normal state as consisting of fermions as well as bosons or pairs of
fermions.  The non-zero $Q$ (finite lifetime or ``resonant") pair
excitations persist below $T_c$ as well, and are not dramatically
affected by the onset of phase coherence.  However, the $Q=0$ pairs
which Bose condense at $T_c$ necessarily acquire an infinite lifetime.
It is this latter effect which is responsible for the signatures of
coherence at $T_c$ in phase insensitive measurements.

Clearly one of the most important questions to be addressed by the
community is to determine whether the pseudogap phase derives from the
superconductivity or from a competing order parameter.  Evidence for the
existence of a quantum critical point would seem consistent with the
latter alternative, although it has been recently
claimed\cite{Loramthisconference} that thermodynamical data may not
provide this support. In favor of the precursor scenario are the
observations that the pseudogap has the same $d$-wave symmetry as the
superconducting order parameter, and that there is no evidence of a
phase transition at $T^*$. In addition, the fact that the high
temperature superconductors seem to belong to a relatively large class
of ``exotic" superconductors as seen by the Uemura plot\cite{Uemura}
suggests that one should focus on generic features of these materials in
order to understand their superconductivity.  Among the most distinctive
generic features of the cuprates and other exotic superconductors is
their short coherence length. It is this latter property which provides
a very likely reason for the failure of BCS theory, and which underpins
the present theoretical approach to the pseudogap phase.

We conjecture that the distinction between the two alternative origins
of the pseudogap may be most clearly evident in the vicinity of the
superconductor-insulator (SI) boundary. It is well established that the
zero temperature excitation gap $\Delta (0)$ is large at this point, but
the competing order parameter scenario\cite{Laughlin}
requires that the zero
temperature order parameter $\Delta_{sc} (0)$ is vanishingly small.  By
contrast, in the absence of quantum fluctuations, the precursor scenario
will have $\Delta_{sc} (0) = \Delta (0) $, which is thus large as well.
In our rendition of this precursor school, superconductivity disappears
because of the localization of $d$-wave pairs\cite{Chen2}.  The
fermionic excitation gap, which is much more robust, persists even after
superconductivity is suppressed, so that the system becomes insulating
once superconductivity is destroyed by doping, by magnetic fields or by
Zn or other impurity substitutions. This ubiquitous SI transition
should hold fundamental clues to the origin of the pseudogap and more
generally to the non-BCS superconductivity of these materials.


This work was supported by the NSF-MRSEC, No.~DMR-9808595 (KL, YK and AI),
by the State of Florida (QC), by the University of Illinois (IK), and
by University of Notre Dame (BJ).



\begin{thebibliography}{10}
\bibitem{companionpaper}
The magnetic field effects are discussed in a companion paper in this
  proceedings.
\end{thebibliography}
\end{document}